\documentclass[twocolumn,epj]{svjour}
\usepackage{latexsym}
\usepackage{graphics}
\usepackage{soul}
\usepackage{graphicx}
\usepackage{color}
\usepackage{dcolumn}
\usepackage{bm}
\usepackage{amsfonts}
\usepackage{amsmath}
\begin{document}
\title{The Palatini formalism of the $f(R,\mathcal{L}_{m},T)$ theory of gravity}
\author{J.G. de Lima Júnior \inst{1}\thanks{\emph{email:} grimario.lima@fis.ufal.br} \inst{} \and P.H.R.S. Moraes \inst{2}\thanks{\emph{email:} moraes.phrs@gmail.com }
 \and E. Brito \inst{3}\thanks{\emph{email:} eliasbaj@ufob.edu.br} \and J.A.S. Fortunato\inst{4}\thanks{\emph{email:} jeferson.fortunato@edu.ufes.br}}
 


%
\institute{Instituto de F\'isica, Universidade Federal do Alagoas (UFAL) - Avenida Lourival Melo Mota S/N, 57072-970, Macei\'o, AL, Brazil \and Laborat\'orio de F\'isica Te\'orica e Computacional (LFTC),
 Universidade Cidade de S\~ao Paulo (UNICID) - Rua Galv\~ao Bueno 868, 01506-000 S\~ao Paulo, Brazil \and Centro de Ci\^encias Exatas e das Tecnologias, Universidade Federal do Oeste da Bahia - Rua Bertioga 892, 47810-059, Barreiras, BA, Brazil \and Centro de Ciências Exatas, Departamento de Física, PPGCosmo, Universidade Federal do Espírito Santo (UFES), Avenida Fernando Ferrari 514, Vitória, ES 29075-910, Brazil} 
\date{Received: date / Revised version: date}
%
\abstract{We present the first formulation of the recently proposed $f(R,\mathcal{L}_m,T)$ theory of gravity within the Palatini formalism, a well-known alternative variational approach where the metric and connection are treated as independent variables. By applying this formalism, we derive a new set of field equations that exhibit, as expected, distinct properties compared to their metric formalism counterparts. We particularly present the Newtonian limit of this formalism, as well as the resulting Friedmann-like equations. We highlight that potential observational signatures may distinguish between the metric and Palatini frameworks. Our results open new pathways for exploring the phenomenology of modified gravity theories and their testability with observational data.}

\maketitle

\section{Introduction}\label{sec:int}

The late-time universe has been undergoing a phase of accelerated expansion. Such a counter-intuitive remarkable dynamical phenomenon was discovered in the late last century from the observation of type Ia supernovae dimmed brightness \cite{riess/1998,perlmutter/1999}. This important discovery was corroborated by other means \cite{weinberg/2013,rubin/2016}. 

Since then, the theoretical description of the cosmic acceleration in accordance with different cosmological probes has been quite a challenge to theoretical physicists. The primary model to attain so is constructed from the inception of the infamous cosmological constant $\Lambda$ in the Einstein's field equations of General Theory of Relativity. However, this approach is mainly haunted by the {\it cosmological constant problem} \cite{weinberg/1989}, so that alternatives are searched.

The main alternatives come from modifications of General Relativity Theory. The geometrical nature of gravity emerges from the universality dictated by the {\it equivalence principle} envisioned by Einstein. General Relativity geometrization of gravity is made in terms of metric and curvature. The metric tensor, however, cannot define curvature by itself so that a connection is required. The connection can have either vanishing or non-vanishing curvature and torsion. In torsion theories, gravity is described by torsion rather than curvature. There are also the non-metricity theories, in which the connection has neither curvature nor torsion. The space-time manifold can be described, therefore, by three different geometrical objects, namely curvature, torsion and non-metricity. This is the so-called geometrical trinity \cite{beltran-jimenez/2019,moraes/2023}. 

The cosmic acceleration has been described in curvature, torsion and non-metricity theories, as one can check, for instance, References \cite{nojiri/2004}, \cite{cai/2016,cardone/2012} and \cite{solanki/2021}, respectively, among many others. Several other applications of the formalisms can be checked \cite{capozziello/2004,perkins/2021,nair/2019,nashed/2021,vasilev/2017,fonseca-neto/1992,de_la_cruz-dombriz/2019,dambrosio/2021}.

It is also possible to insert material terms in the gravitational action. In the curvature theories, this yields the $f(R,\mathcal{L}_{m})$ \cite{harko/2010}, $f(R,T)$ \cite{harko/2011} and $f(R,\mathcal{L}_{m},T)$ \cite{haghani/2021} theories, in which $R$, $\mathcal{L}_m$ and $T$ are, respectively, the Ricci scalar, matter lagrangian density and trace of the energy-momentum tensor. Note that extend the $f(R)$ theory to encompass $\mathcal{L}_{m}$ and/or $T$-terms in the gravitational action is motivated by difficulties $f(R)$ gravity presents in predicting the transition from a decelerated to an accelerated universe expansion \cite{amendola/2006,amendola/2006b,amendola/2007}, getting rid of singularity problems in dark energy models \cite{frolov/2008} and passing solar system tests \cite{chiba/2003,chiba/2007,erickcek/2006,nojiri/2008,capozziello/2007,capozziello/2008,olmo/2007}. Even relativistic stars cannot be present in $f(R)$ theories according to \cite{kobayashi/2008}. For torsion and non-metricity theories with material terms in the action, check \cite{harko/2014b,junior/2016,xu/2019,najera/2022,arora/2020,myrzakulov/2023}.

From now on, let us concentrate on the curvature theories only. When deriving the field equations of a given theory from the variational principle applied to the theory's action, the variations are taken with respect to the metric tensor coefficients, while the connections are assumed to be Christoffel symbols defined in terms of the metric. This is usually referred to as the {\it metric formalism}. An alternative procedure, originally considered by Einstein himself, is to treat both metric and connections as independent variables and perform the variations with respect to both of them. This is usually referred to as the {\it Palatini formalism}.

Our intent in the present paper is to obtain the Palatini formalism version of the aforementioned recently proposed $f(R,\mathcal{L}_{m},T)$ theory of gravity. The theory was presented in the metric formalism in \cite{haghani/2021} and already presents some interesting applications in the literature. For instance, Zubair et al. gave a perturbative analysis of $f(R,\mathcal{L}_{m},T)$ models in \cite{zubair/2023}. Hydrostatic equilibrium configurations of compact stars in the theory were presented in \cite{fortunato/2024,mota/2024}. Wormhole solutions can be seen in \cite{moraes/2024}. Finally, the authors of the theory themselves confront the theoretical predictions of $f(R,\mathcal{L}_{m},T)$ models with cosmological data in \cite{haghani/2021}.

Considering the Palatini formalism in modified gravity theories is crucial for advancing our understanding of gravitational phenomena in non-classical regimes. To treat the metric and the connection as independent variables offers a more flexible framework for theoretical modeling. This method not only can provide alternative insights into the structure of space-time itself but also facilitates the exploration of solutions that address fundamental questions about the nature of gravity and its interaction with other fundamental forces \cite{teruel/2013}. Consequently, adopting the Palatini formalism can yield significant insights that help overcome the limitations of conventional gravity theories. The Palatini formalism has been exhaustively used in the $f(R)$ theory, as one can check \cite{li/2007,stachowski/2017,bamba/2010,fay/2007,dominguez/2004,kucukakca/2012,mohaved/2007,gogoi/2022}, for instance, and here, it will be obtained, for the first time in the literature, for the $f(R,\mathcal{L}_{m},T)$ gravity.

The present article is structured as follows. In Section \ref{sec:br}, we present a brief review of the $f(R,\mathcal{L}_{m},T)$ gravity theory in the metric formalism. In Section \ref{sec:pf}, we consider the formulation ``à la Palatini'' for the $f(R,\mathcal{L}_m,T)$ gravity theory. In Section \ref{sec:nl} we investigate the Newtonian limit of the $f(R,\mathcal{L}_m,T)$ theory in the Palatini formalism. In Section \ref{sec:cos} we derive the Friedmann-like equations in the formalism. Our final remarks are presented in Section \ref{sec:fr}.

\section{A brief review of the $f(R,\mathcal{L}_{m},T)$ theory of gravity in the metric formalism}\label{sec:br}

As originally proposed, the $f(R,\mathcal{L}_{m},T)$ gravity theory starts from the action \cite{haghani/2021}

\begin{equation}\label{br1}
    S=\int d^4x\sqrt{-g}\left[\frac{f(R,\mathcal{L}_{m},T)}{16\pi}+\mathcal{L}_{m}\right].    
\end{equation}
In \eqref{br1}, $g$ is the determinant of the metric tensor and throughout the paper, natural units will be assumed. From the action above, it is possible to see that the functional form of the function $f$ may retain generic terms of non-minimal coupling between geometry and matter. 

By varying \eqref{br1} with respect to the metric yields the field equations of the theory in the metric formalism, namely

\begin{eqnarray}\label{br2}
    &&(R_{\mu\nu}+g_{\mu\nu}\Box-\nabla_\mu\nabla_\nu)f_R-\frac{1}{2}(f-2f_m\mathcal{L}_{m})g_{\mu\nu}=\nonumber\\
    &&(8\pi+f_m)T_{\mu\nu}+f_T\tau_{\mu\nu},
\end{eqnarray}
for which $R_{\mu\nu}$ is the Ricci tensor, $g_{\mu\nu}$ is the metric tensor, $f_R\equiv\partial f/\partial R$, $f_m\equiv f_{\mathcal{L}_{m}}/2+f_T$, $f_{\mathcal{L}_m}\equiv\partial f/\partial \mathcal{L}_{m}$, $f_T\equiv\partial f/\partial T$, 
\begin{equation} \label{tEM_1}
T_{\mu\nu}\equiv g_{\mu\nu}\mathcal{L}_{m} -\frac{2\partial \mathcal{L}_{m}}{\partial g^{\mu\nu}}    
\end{equation}
is the energy-momentum tensor and 

\begin{equation}\label{br2.1}
   \tau_{\mu\nu}\equiv2g^{\alpha\beta}\frac{\partial^2\mathcal{L}_{m}}{\partial g^{\mu\nu}\partial g^{\alpha\beta}}.
\end{equation}

Finally, the equation for the covariant derivative of the energy-momentum tensor reads

\begin{eqnarray}\label{br3}
    (8\pi+f_m)\nabla^\mu T_{\mu\nu}&=&\nabla_\nu(f_m\mathcal{L}_{m})-T_{\mu\nu}\nabla^\mu f_m-\nabla^\mu(f_T\tau_{\mu\nu})\nonumber\\
    &-&\frac{1}{2}(f_{\mathcal{L}_{m}}\nabla_\nu \mathcal{L}_{m}+f_T\nabla_\nu T).
\end{eqnarray}

\section{$f(R,\mathcal{L}_{m},T)$ in Palatini formalism}\label{sec:pf}

In this section we will consider the formulation ``à la Palatini'' for the $f(R,\mathcal{L}_{m},T)$ theory. In this approach, unlike what occurs in the metric formalism, the connection $\Gamma^{\alpha}_{\mu\nu}$ is independent of the metric $g_{\mu\nu}$. In this case, we will adopt a Ricci tensor $\mathcal{R}_{\mu\nu}$, which is formed from the independent connection and the metric. Because of this we also adopt a Ricci scalar $\mathcal{R}=g^{\mu\nu}\mathcal{R}_{\mu\nu}$.

A key advantage of working with the Palatini formalism is that, unlike the metric formalism, it avoids instabilities since it yields second-order differential field equations for the metric. This feature prevents the emergence of ghost fields, as predicted by Ostrogradski's theorem \cite{Ostrogradsky:1850fid}, which typically arise in higher-order theories. Another extremely important quality is that, also due to the natural structure of this formalism, the equations for the connection components are bonds, and not dynamic equations, which, in other words, means that they do not introduce new dynamic fields.

\subsection{Field equations from the variation with respect to the metric}

Taking the variation of Eq.\eqref{br1} with respect to the metric yields the field equations
\begin{eqnarray} \label{eqcg} \nonumber
    &&f_{\mathcal{R}}\mathcal{R}_{\mu\nu} -\frac{f}{2}g_{\mu\nu} =  \\ 
    &&\kappa^{2}T_{\mu\nu} -\frac{f_{\mathcal{L}_m}}{2}\left( g_{\mu\nu}\mathcal{L}_{m}-T_{\mu\nu}\right) -f_{T}\left( T_{\mu\nu} +\Theta_{\mu\nu}\right) .
\end{eqnarray}
Note also that the variation of $T_{\mu\nu}$ with respect to the metric tensor has the form
\begin{equation}
    \frac{\delta(g^{\alpha \beta}T_{\alpha\beta})}{\delta g^{\mu\nu}}=T_{\mu\nu}+\Theta_{\mu\nu}
\end{equation}
where, we are defining
\begin{equation} \label{tTH}
    \Theta_{\mu\nu} \equiv g^{\alpha\beta}\frac{\delta T_{\alpha\beta}}{\delta g^{\mu\nu}} .
\end{equation}
The trace of Eq. \eqref{eqcg} gives
\begin{equation}\label{teqcg}
        f_{\mathcal{R}}\mathcal{R} -2f = \kappa^{2}T -f_{\mathcal{L}_m}\left(2\mathcal{L}_{m} -\frac{T}{2}\right) -f_{T}\left( T +\Theta\right).
\end{equation}

With Eqs. \eqref{eqcg} and \eqref{teqcg}, we can write the Einstein tensor in terms of the variation with respect to the metric only:
\begin{eqnarray}
    \label{Etg} \nonumber
    \mathcal{G}_{\mu\nu} &=&\frac{1}{f_{\mathcal{R}}}\left\{\kappa^{2}\left(T_{\mu\nu} -\frac{g_{\mu\nu}}{2}T\right)-\frac{f_{\mathcal{L}_{m}}}{2} \biggr[ g_{\mu\nu}\mathcal{L}_{m}-T_{\mu\nu} \right. \\ \nonumber
 && \left. \left. -g_{\mu\nu}\left(2\mathcal{L}_{m}-\frac{T}{2} \right)\right] - f_{T} \biggr[(T_{\mu\nu}+\Theta_{\mu\nu}) \right. \\
 && \left. -\frac{g_{\mu\nu}}{2}\left(T+\Theta \right) \biggr] - \frac{g_{\mu\nu}f}{2} \right\} .
\end{eqnarray}

\subsection{Field equations from the variation with respect to the connection}

Now, considering the variation of the gravitational action with respect to the connection, we obtain
\begin{equation} \label{eqcc}
\nabla_{\lambda}\left(\sqrt{-g}f_{\mathcal{R}}g^{\mu\nu}\right)=0 .
\end{equation}
Note that the field equations obtained above are identical to those found in both the $f(R)$ and $f(R,T)$ theories \cite{Wu:2018idg}. This makes it convenient to introduce a new second-order tensor $h_{\mu\nu}$, which we call ``auxiliary metric'', where
\begin{equation} \label{auxh}
h_{\mu\nu}\equiv f_{\mathcal{R}}g_{\mu\nu} \ ,
\end{equation}
such that
\begin{equation}
    \nabla_{\lambda}\left(\sqrt{-h}h^{\mu\nu}\right)=0  .
\end{equation} This allows us to directly solve Eq.\eqref{eqcc}, whose solution is the Levi-Civita connection. Therefore, the connection in terms of the new metric reads

\begin{equation}
\Gamma^{\lambda}_{\mu\nu}=\frac{1}{2}h^{\lambda\sigma}\left(\partial_{\mu} h_{\nu\sigma}+\partial_{\nu}h_{\mu\sigma}-\partial_{\sigma}h_{\mu\nu} \right) .
\end{equation}

The Ricci tensor transforms as \cite{Sotiriou:2008rp}
\begin{eqnarray} \label{Rtc} \nonumber
\mathcal{R}_{\mu\nu}&=&R_{\mu\nu}+\frac{1}{f_{\mathcal{R}}}\biggr[\frac{3}{2}\frac{1}{f_{\mathcal{R}}}\nabla_{\mu}f_{\mathcal{R}}\nabla_{\nu}f_{\mathcal{R}} \\
&&-\left(\nabla_{\mu}\nabla_{\nu}+\frac{1}{2}g_{
\mu\nu}\Box \right)f_{\mathcal{R}}\biggr],
\end{eqnarray}
and consequently the Ricci Scalar transforms as
\begin{equation} \label{Rsc}
 \mathcal{R}=R+\frac{3}{f_{\mathcal{R}}}\left(\frac{1}{2f_{\mathcal{R}}}\nabla_{\mu}f_{\mathcal{R}}\nabla^{\mu}f_{\mathcal{R}}-\Box f_{\mathcal{R}}\right) .
\end{equation}

From Eqs.\eqref{Rtc} and \eqref{Rsc}, we can write the Einstein tensor as 

\begin{eqnarray}
    \label{Etc} \nonumber
    \mathcal{G}_{\mu\nu} &=& G_{\mu\nu} + \frac{1}{ f_{\mathcal{R}}}\left[\frac{3}{2 f_{\mathcal{R}}}\left(\nabla_{\mu}  f_{\mathcal{R}} \nabla_{\nu} f_{\mathcal{R}} -\frac{g_{\mu\nu}}{2} \left( \nabla f_{\mathcal{R}} \right)^{2}\right) \right. \\
 &&  +  \left(g_{\mu\nu} \Box -\nabla_{\mu}\nabla_{\nu}\right)f_{\mathcal{R}} \biggr] .
\end{eqnarray}

Finally, with Eqs.\eqref{Etg} and \eqref{Etc}, we can write the Einstein tensor in its complete form for the $f(\mathcal{R}, \mathcal{L}_{m}, T)$ theory of gravity:
\begin{eqnarray}
     \label{EtP} \nonumber
    G_{\mu\nu}&=& \frac{1}{ f_{\mathcal{R}}} \left\{ -\left[\left(g_{\mu\nu} \Box -\nabla_{\mu}\nabla_{\nu}\right)f_{\mathcal{R}} +\frac{3}{2 f_{\mathcal{R}}}\biggl(\nabla_{\mu}  f_{\mathcal{R}} \nabla_{\nu} f_{\mathcal{R}}\right. \right. \\ \nonumber
 && \left. \left. \left. -\frac{g_{\mu\nu}}{2} \left( \nabla f_{\mathcal{R}} \right)^{2}\right) \right] + \kappa^{2}\left(T_{\mu\nu} -\frac{g_{\mu\nu}}{2}T\right) \right. \\  \nonumber
 && \left.  -\frac{f_{\mathcal{L}_{m}}}{2}\left[g_{\mu\nu}\mathcal{L}_{m}-T_{\mu\nu}-g_{\mu\nu}\left(2\mathcal{L}_{m}-\frac{T}{2} \right)\right] \right. \\
 && \left. - f_{T}\left[(T_{\mu\nu}+\Theta_{\mu\nu})-\frac{g_{\mu\nu}}{2}\left(T+\Theta \right) \right] - \frac{g_{\mu\nu}f}{2} \right\} .
\end{eqnarray}

Note that the inclusion of terms referring to the matter Lagrangian density and trace of the energy-momentum tensor in the gravitational Lagrangian is responsible only for a modification in the field equations referring to the metric, whereas for the connection they are the same as what is obtained when we treat the case of a theory in which the gravitational Lagrangian depends only on $R$.

\section{The Newtonian limit}\label{sec:nl}

Let us investigate the Newtonian limit for the $f(R, \mathcal{L}_{m}, T)$ theory in the Palatini formalism. To do this, let us consider our metric $g_{\mu\nu}$ to be the Minkowski metric $\eta_{\mu\nu}$ plus a perturbation, i.e.,

\begin{equation} \label{pertubation}
    g_{\mu\nu}=\eta_{\mu\nu} +\gamma_{\mu\nu}
\end{equation}
where $\gamma_{\mu\nu}\ll 1$.

In this context, we also assume the auxiliary metric \eqref{auxh} to be nearly flat, so that
\begin{equation} \label{pertubation2}
    h_{\mu\nu}=f_{\mathcal{R}}g_{\mu\nu}=\eta_{\mu\nu} +\Tilde{\gamma}_{\mu\nu} ,
\end{equation}
where $\Tilde{\gamma}_{\mu\nu}\ll 1$ is of the same order as $\gamma_{\mu\nu}$. Hence, $f_{\mathcal{R}}\approx 1$. Using Eqs, \eqref{pertubation} and \eqref{pertubation2} we obtain
\begin{equation}
    (f_{\mathcal{R}}-1)\eta_{\mu\nu}=\Tilde{\gamma}_{\mu\nu} -f_{\mathcal{R}}\gamma_{\mu\nu} .
\end{equation}
If we take $f_{\mathcal{R}}=e^{2W}$ and expand it at first order as $f_{\mathcal{R}}=1+2W$, then the above equation shows that 

\begin{equation}
  W=\frac{\Tilde{\gamma}-\gamma}{2(4+\gamma)}.
\end{equation} 

where $\Tilde{\gamma}\equiv \eta^{\mu\nu}\Tilde{\gamma}_{\mu\nu}$ and $\gamma\equiv \eta^{\mu\nu}\gamma_{\mu\nu}$. Thus, $W\sim O(\gamma)\sim O(\gamma_{\mu\nu})$.

Now, let us write the $g$-frame Ricci tensor:

\begin{equation}
    R_{\mu\nu} = \frac{1}{2} \left( \partial^{\lambda} \partial_{\mu} \gamma_{\nu \lambda} + \partial^{\lambda} \partial_{\nu} \gamma_{\mu \lambda} - \partial^{2} \gamma_{\mu \nu} - \partial_{\mu \nu} \gamma \right).
\end{equation}

The Ricci tensor expression \eqref{Rtc} can be rewritten as
\begin{equation} \label{Rtc_Nl}
\mathcal{R}_{\mu\nu}= -\frac{1}{2}\partial^{2}(\gamma_{\mu\nu} +2\eta_{\mu\nu}W ) -2\partial_{\mu\nu}W.
\end{equation}

The field equations \eqref{eqcg} under the Newtonian limit now become

\begin{eqnarray} \label{eqcg_Nl} \nonumber
    &&-\frac{1}{2}\partial^{2}\Tilde{\gamma}_{\mu\nu} -2\partial_{\mu\nu}w -\frac{f}{2}\eta_{\mu\nu} =  \\ 
    &&\kappa^{2}T_{\mu\nu} -\frac{f_{\mathcal{L}}}{2}\left( \eta_{\mu\nu}\mathcal{L}_{m}-T_{\mu\nu}\right) -f_{T}\left( T_{\mu\nu} +\Theta_{\mu\nu}\right) .
\end{eqnarray}

Considering the case of a perfect fluid characterized only by its energy density $\rho$ and isotropic pressure $p$, the energy-momentum tensor is given by
\begin{equation} \label{tEM_2}
    T_{\mu\nu} = (\rho +p)u_{\mu}u_{\nu} +p g_{\mu\nu}.
\end{equation}
In the comoving frame, the four-velocity components are $u^{\mu}=(-1, 0, 0, 0)$, such that the four-velocity $u^{\mu}$ satisfies the normalization condition $u^{\mu}u_{\mu}=-1$. In this frame, the components of the energy-momentum tensor become $T_{\mu\nu}=\text{diag}(-\rho , p, p, p)$, such its trace reads

\begin{equation}
    T= -\rho +3p.
\end{equation}

Making use of the definition of the $\Theta_{\mu\nu}$ tensor and Eq.\eqref{tEM_1}, we can write this tensor for a perfect fluid:
\begin{equation}
   \Theta_{\mu\nu} = g_{\mu\nu}\mathcal{L}_{m} -2T_{\mu\nu} -2g^{\alpha\beta}\frac{\partial^{2}\mathcal{L}_{m}}{\partial g^{\mu\nu}\partial g^{\alpha\beta}},
\end{equation}
such that considering $\mathcal{L}_{m}=p$, we have
\begin{equation}
    \Theta_{\mu\nu}=g_{\mu\nu} -2 T_{\mu\nu}.
\end{equation}

Now we can write Eq.\eqref{eqcg_Nl} as
\begin{eqnarray} \nonumber
&&-\frac{1}{2}\partial^{2}\Tilde{\gamma}_{\mu\nu} -2\partial_{\mu\nu}w -\frac{f}{2}\eta_{\mu\nu}= \\
&&\rho \left(-\kappa^{2} -f_{T} -\frac{f_{\mathcal{L}}}{2} \right)u_{\mu}u_{\nu} -p\left( \frac{f_{\mathcal{L}}}{2} +f_{T} \right) \eta_{\mu\nu}.
\end{eqnarray}

Taking the $00$ component of the previous expression, which governs the weak field dynamics, that is, in the Newtonian limit ($p \to 0$), space-time is assumed to be static and the time derivatives are zero, which leads us to the following expression:

\begin{equation}
    -\frac{1}{2}\Vec{\nabla}^{2}\Tilde{\gamma}_{00}= \rho (\kappa^{2} -f_{T} +f_{\mathcal{L}} ) -\frac{f}{2},
\end{equation}
which is the Poisson's equation in the Palatini formulation of the $f(\mathcal{R},\mathcal{L}_{m},T)$ theory of gravity.

\section{Friedmann-like equations}\label{sec:cos}

In this section we will construct the Friedmann-like equations for the $f(R,\mathcal{L}_m,T)$ theory in the Palatini formalism. For particular choices of the $f(R,\mathcal{L}_m,T)$ function, those can further be confronted with cosmological data, with the purpose of verifying if the Palatini formulation of the $f(R,\mathcal{L}_m,T)$ gravity can explain the cosmic acceleration (naturally, with no need for the cosmological constant). We start from the Friedmann-Lemâitre-Robertson-Walker metric:

\begin{equation} \label{frlw}
    ds^{2}= -dt^{2} + a(t)^{2}(dx^{2} +dy^{2} +dz^{2})
\end{equation}
where $a(t)$ is the scale factor, such that $H=\Dot{a}/a$ is the Hubble parameter. 

From the $00$ component of Eq.\eqref{EtP}, considering the metric above and the energy-momentum tensor \eqref{tEM_2}, we obtain the first modified Friedmann equation as
\begin{eqnarray} \label{Friedmann_00_1} \nonumber
    3H^{2}&=&\frac{1}{2f_{\mathcal{R}}}\left[\kappa^{2}(\rho +3p) -\frac{f_{m}}{2}(3\rho -13 p+8\mathcal{L}_{M}) \right. \\
 && \left. +f -\frac{3(\Dot{f}_{\mathcal{R}})^{2}}{2 f_{\mathcal{R}}} -6H\Dot{f}_{\mathcal{R}}\right],
\end{eqnarray}
or even, in a more compact way
\begin{equation} \label{Friedmann_00_2}
    \left(H +\frac{\Dot{f}_{\mathcal{R}}}{2f_{\mathcal{R}}} \right)^{2} =\frac{\kappa^{2}(\rho +3p)-\frac{f_{m}}{2}(3\rho -13p +8\mathcal{L}_{M})+f}{6f_{\mathcal{R}}}.
\end{equation}

Similarly, we can obtain the second modified Friedmann equation from the $ii$ components:
\begin{eqnarray} \label{Friedmann_ii} \nonumber
    2\Dot{H}+3H^{2} &=& -\frac{1}{2f_{\mathcal{R}}}\left[\kappa^{2}(\rho -p) -\frac{f_{m}}{2}(3\rho -5p +8\mathcal{L}_{M}) \right. \\
 && \left.  -f -\frac{3(\Dot{f}_{\mathcal{R}})^{2}}{2 f_{\mathcal{R}}} -2\Ddot{f} -4H\Dot{f}_{\mathcal{R}}\right].
\end{eqnarray}

An equation for the evolution of the Hubble function comes from the elimination of the term $3H^{2}$ in the Friedmann equations:
\begin{eqnarray} \nonumber
\Dot{H}&=&-\frac{1}{2f_{\mathcal{R}}}\left[\kappa^{2}(\rho +p) -f_{m}\left(4\mathcal{L}_{M} +\frac{3}{2}\left(\rho +p \right) \right) \right. \\
 && \left. -\frac{3(\Dot{f}_{\mathcal{R}})^{2}}{2} +\Ddot{f} -H\Dot{f}_{\mathcal{R}}\right] .
\end{eqnarray}


\section{Final remarks}\label{sec:fr}

In the present article, we have obtained, for the first time in the literature, the Palatini formalism of the recently proposed $f(R,\mathcal{L}_m,T)$ gravity. By employing the variational approach, we derived field equations that exhibit distinct properties compared to their metric formalism counterparts. With our results, we offer a new set of equations to be applied to different systems and regimes, that will ultimately verify the viability of the theory. 

We have particularly obtained the Poisson's equation in the formalism as well as the Friedmann equations. Together, and in possession of the referred observational data, those can put stringent constraints to the functional form of the $f(R,\mathcal{L}_m,T)$ function.

The Palatini formulation of this theory can also bring some valuable insights on the geometry-matter non-minimal coupling allowed within the theory. Theories with non-minimal coupling between geometry and matter have intensified their appearance in the field of modified gravity and offer a more fundamental relation between matter and space-time curvature, with an optimistic example being the recently proposed Entangled Relativity \cite{minazzoli/2018}.

\section*{Data availability} There are no new data associated with this article.

\section*{Acknowledgements}

JGLJ thanks CAPES for financial support. PHRSM would like to thank CNPq (Conselho Nacional de Desenvolvimento Cient\'ifico e Tecnol\'ogico) for partial financial support under grant No. 310366/2023-2. JASF thanks FAPES for financial support.

\bibliographystyle{unsrt}{99}
\bibliography{Bib.bib}

\end{document}